\documentclass[prd,aps,reprint,onecolumn,10pt,showpacs,preprintnumbers,amsmath,amssymb,floatfix,superscriptaddress]{revtex4}

\usepackage{graphicx} 
\usepackage{dcolumn}  
\usepackage{bm}       
\usepackage{hyperref}
\usepackage{color}
\usepackage{epsfig}
\usepackage{float}

\newcommand{\beq}{\begin{equation}}
\newcommand{\eeq}{\end{equation}}
\newcommand{\ds}{\displaystyle}

\begin{document}

\preprint{Preprint}

\title[A fully relativistic radial fall]{A fully relativistic radial fall}

\author{Alessandro D.A.M. Spallicci
}
\affiliation{Universit\'e d'Orl\'eans\\ 
Observatoire des Sciences de l'Univers en r\'egion Centre, OSUC - UMS 3116 \\
Laboratoire de Physique et Chimie de l'Environnement et de l'Espace, LPC2E - UMR CNRS 7328\\
3A Avenue de la Recherche Scientifique, 45071 Orl\'eans, France}
\author{Patxi Ritter
}
\affiliation{Universit\'e d'Orl\'eans\\ 
Observatoire des Sciences de l'Univers en r\'egion Centre, OSUC - UMS 3116 \\
Laboratoire de Physique et Chimie de l'Environnement et de l'Espace, LPC2E - UMR CNRS 7328\\
3A Avenue de la Recherche Scientifique, 45071 Orl\'eans, France}
\affiliation{Universit\'e d'Orl\'eans\\
Math\'ematiques, Analyse, Probabilit\'es, Mod\`elisation - Orl\'eans, MAPMO - UMR CNRS 7349\\
Rue de Chartres, 45067 Orl\'eans, France} 

\date{21 July 2014}
\begin{abstract}
Radial fall has historically played a momentous role. It is one of the most classical problems, the solutions of which represent the level of understanding of gravitation in a given epoch. A {\it gedankenexperiment} in a modern frame is given by a small body, like a compact star or a solar mass black hole, captured by a supermassive black hole. The mass of the small body itself and the emission of gravitational radiation cause the departure from the geodesic path due to the back-action, that is the self-force. 
For radial fall, as any other non-adiabatic motion, the instantaneous identity of the radiated energy and the loss of orbital energy cannot be imposed and provide the perturbed trajectory. In the first part of this letter, we present the effects due to the self-force computed on the geodesic trajectory in the background field. Compared to the latter trajectory, in the Regge-Wheeler, harmonic and all others smoothly related gauges, a far observer concludes that the self-force pushes inward (not outward) the falling body, with a strength proportional to the mass of the small body for a
given large mass; further, the same observer notes an higher value of the maximal coordinate velocity, this value being reached earlier on during infall.  In the second part of this letter, we implement a self-consistent approach for which the trajectory is iteratively corrected by the self-force, this time computed on osculating geodesics. Finally, we compare the motion driven by the self-force without and with self-consistent orbital evolution. Subtle differences are noticeable, even if self-force effects have hardly the time to accumulate in such a short orbit. 
\end{abstract}


\maketitle

\section{The epistemology of free fall}

Free fall has been a source of inspiration in many instances. A popular misconception identifies the equivalence principle (EP) or the uniqueness of free fall (UFF) with the ill-founded concept that lighter and heavier bodies accelerate equally.  
The simplest formulation of the EP affirms that inertial mass and gravitational mass are the same thing. Then, the gravitational force is proportional to inertial mass, and the proportionality is independent of the kind of matter. For simplicity, the coefficient of proportionality is fixed to unity. In a uniform gravitational field the UFF implies that objects of same mass value but different composition fall with the same acceleration. 

Instead, for objects of different mass value, there is no EP or UFF dictating the same acceleration. Indeed, two (small at leisure) bodies of different mass value $m_1$ and $m_2$ do not accelerate equally in a field of a large mass $M$. In Newtonian mechanics, the displacement of the centre of mass (from the centre of $M$ to a new location dependent upon the small mass $m$) corresponds to a self-force taking into account the finitude of the small mass. To masses in  circular motion of radius $r_c$, the Newtonian self-force induces   faster or slower angular speeds than the nominal value $\omega = \sqrt{GM/r_c^3}$, $G$ being the gravitation constant. At first order, the difference amounts to $\pm n m/M$, where the sign and the value of $n$ depend upon the chosen coordinate system \cite{depo04,de08,de11}. 

The preceding has been extended to radial fall \cite{sp11}, as follows. We set the origin of the reference frame at the centre of mass of the system $r_M M = r_m m$. The acceleration for $m$ is then given by ${\ddot r}_m = - GM (1 - 2m/M)/r_m^2 $, and we infer that heavier bodies fall slower. Instead, for $h$ being the altitude of $m$ and $r_\oplus$ the Earth radius, ${\ddot h}= - GM (1 + m/M)/(r_\oplus + h)^2 $, and we infer that heavier bodies fall faster! Thus, the sign and amplitude dependence of the self-force on the chosen coordinates is a manifestation of gauge in Newtonian mechanics. 

Equating the centre of mass of the system to the centre of mass of the large body deletes any self-force effect, and this translation can always be arranged. But the translation of the centre of mass of the two-body system is an operation that patently differs according to the mass value of the small body. Therefore, for two contemporaneously infalling bodies of different mass value, there is no translation possible that cancels self-force effects for both of them, and the concepts of uniqueness and universality of free fall lose any significance in this respect. 

Ironically, if the two bodies are not only of different mass value, but also of different composition, a violation of the EP may counteract  the acceleration difference due to the different mass values, and provide a null result.

Anyway, on the Earth and in its surroundings, the Newtonian self-force appears negligible ($\approx 10^{-25}$~m s$^{-2}$ for each kg), given the current technology for the upcoming measurements \cite{tomelero12}.  

\section{Coordinate effects for geodesic trajectories}

With the advent of general relativity, the Earth - the reference for several {\it gedankexperimente} - was replaced by the black hole 
\footnote{The first solution of the Einstein equations was proposed in January 1916 by Schwarzschild \cite{sc16}, and independently in May by Droste \cite{dr16a, dr16b}, the SD metric.}. 
Observers in general relativity describe the phenomena
according to their chosen coordinate system or gauge;
thus the importance of finding and dealing with gauge invariant
quantities. Nevertheless, non-invariant quantities
are also meaningful, when the associated coordinate system
is clearly spelled out. A well-known coordinate dependent
description dictates that a particle never crosses
the horizon of a black hole for a far observer, while an
observer at the particle crosses the horizon.

Despite the mathematical simplicity, a controversy on a falling mass into Schwarzschild-Droste (SD) geometry \cite{ei87,sp11} took place from 1916, before decaying after the '80s.   
The debate, often reignited for the oblivion of earlier literature,  	
reflected the unawareness of coordinate choice, while the debaters showed affection to a frame considered more `physical'. 
The question was: is there an effect of repulsion? Or does the particle speed reach a maximal velocity and then slows down? 
Oblivion of earlier literature has
caused the concepts of repulsion and critical velocity to be
'rediscovered' again and again. Since critical velocity is advocated in high energy astrophysics, we explain how the concept originated. 

Four types of measurements (-, +, +, + signature) might be envisaged: affected by gravity $dl$, $d\tau$, or not (a far observer) $dr$, $dt$ 
($dl = (1 - r_{\rm g}/r)^{-1/2} dr$ and $d\tau = - (1 - r_{\rm g}/r)^{1/2} dt$, $r_{\rm g} = 2GM/c^2$ the gravitational radius).
Following Thirring \cite{th61} and Sexl \cite{se67}, Cavalleri and Spinelli \cite{casp73, sp89} define the velocities and accelerations as renormalised, {\it id est}, proper ($dl/d\tau$ and $d^2l/d\tau^2$), non-renormalised ($dr/dt$ and $d^2r/dt^2$), and semi-renormalised ($dl/dt$ and $d^2l/dt^2$ or $dr/d\tau$ and $d^2r/d\tau^2$).  
The quantities $dl/dt$ and $d^2l/dt^2$ are measured by a far observer who uses his own clock, but equivalently measures distances with a meter stick placed at the particle,  or by comparing the position of objects which are local to the particle,  
or by integrating the echo times of signals reflected by the particle, as suggested by Jaffe and Shapiro \cite{jash72, jash73}. 

For a null velocity at infinity, Droste \cite{dr16a, dr16b} already shows that $d^2l/dt^2$ turns positive for either $ r \leq 4GM/c^2 $ or else $| dl/dt| \geq c \sqrt{1/2}\sqrt{1- r_{\rm g}/r}$; sharing the same conviction, von Laue \cite{vl21}, Bauer \cite{ba22}, McVittie \cite{mv56}, M{\o}ller \cite{mo60}, and the proposers of a satellite experiment \cite{kuza10}. Still for null velocity at infinity, Droste \cite{dr16a} finds that the acceleration $d^2 r/dt^2$ is positive for either $ r \leq 6GM/c^2 $ or else $|dr/dt| \geq c\sqrt{1/3}(1 - r_{\rm g}/r)$; on the same stand, Hilbert \cite{hi17}, Page \cite{pa20a}, Eddington \cite{ed20}, Treder \cite{tr72, trfr75}, Carmeli \cite{ca72, ca82}.  

von Laue himself \cite{vl26} and Drumaux \cite{dr36} propose first the proper acceleration $d^2l/d\tau^2$ for which no repulsion occurs; on this line opposing any repulsion, Landau and Lifshits \cite{lali41}, von Rabe \cite{vr47}, Whittaker \cite{wh53}, Zel'dovich and Novikov \cite{zeno67}, and others \cite{ma73, ar81}. 
For other views on the matter see \cite{sr66,ba73,ri79,shte83,frno98,bo07,bo13,mu08,wi13}. 

Geodesic motion and particle speed at the horizon were also discussed  \cite{ta81,dapa82,ja73,ja77,casp77,casp78,crte02,mi02}.   McGruder III points out that repulsion is depending upon the relation between radial  and transverse velocities \cite{mc82}; Kerr and Kerr-Newman geometries were also analysed \cite{krba85}.

From Blinnikov {\it et al.} \cite{blokvy01, blokvy03breve}, for a null starting value at $r_0$, we draw that the non-renormalised velocity $|dr/dt| = - g_{00}c[ 1 - g_{00}/g_{00}(r_0)]^{1/2}$ reaches the maximal value of $|dr/dt|_{\rm max} = 2 c(r_0 - r_{\rm g})/(3 \sqrt{3} r_0)$ at $r_{\rm max} = 3 r_0 r_{\rm g}/(r_0 + 2 r_{\rm g})$, for $g_{00}$ being the SD metric time coefficient. For the coordinate acceleration being null at $r_{\rm max}$, the local velocity is $|dl/d\tau| =  - g_{00}^{-1}dr/dt$ is $ c(r_0 - r_{\rm g})/[(r_0 + r_{\rm g})\sqrt{3}]$. In far field, the preceding extends to other asymptotically flat coordinates. 

Maximal velocities are discussed in high energy astrophysics \cite{chma05,ma05}. Therein it is claimed that tidal forces exhibit a peculiar behaviour due to the maximal velocity. A mass of finite size, falling with a speed lower than the maximal value towards a Kerr black hole, is stretched along the longitudinal axis, supposed coincident with the rotation axis, and compressed along the transversal axes, just like in Newtonian physics. But  
for a higher speed, the same mass would conversely be compressed along the longitudinal axis and stretched on the transversal ones. 
The ultra-relativistic particles near a gravitationally
collapsed system or in the accretion process are thus tidally decelerated in a cone around the rotation
axis of the collapsed system, and correspond to jets from
neutron stars and X-ray binaries. Meanwhile, the ultra-relativistic
particles that result from tidal acceleration outside the
cone transfer their tidal energy to the ambient
particles, thus inducing extremely energetic cosmic rays. 

Maximal velocity has been considered for stars orbiting black holes \cite{kota06}. The stars are disrupted by tidal forces and produce luminous flares, followed by a declining phase, as X-ray observations suggest. The authors speculate that the disruption of a rapidly rotating star due to a velocity dependent tidal force is quite different from that of a non-relativistic star.

These papers refer to Fermi coordinates \cite{fe22a,fe22b,fe22c}, for which the maximal velocity is $c/\sqrt{2}$. The Fermi coordinates were originally conceived for a small region relative to the length scale. Therefore, the implications that these authors draw on the observations for other observers are not evident \cite{chma06,dean12}.      

Concerning photons, for a far observer, they indeed ``slow down'' when approaching a mass, forming the radio echo delay \cite{sh64,jash72,jash73}.

\section{Impact of the self-force on the trajectory} 

In the '70s, Zerilli computes the gravitational radiation emitted during infall \cite{ze70a,ze70b,ze70c}. Forty years later, back-action - without orbital evolution - was partially analysed only in two works \cite{lo00,balo02}, and with contrasting predictions (the former suggests the back-action to be repulsive for some modes and attractive for others, conversely to the second which attributes always an attractive feature). Though we present herein only new results, we comment the previous 
literature \footnote{We differ with the findings in \cite{lo00}: i) In the upper panel of Fig. 2 in \cite{lo00}, there is a dependence of the sign of $a_{\rm self}$ on $\ell$ (an attractive contribution for $\ell\leq 3$ and repulsive for $\ell > 3$), and the presence of the $\ell = 1$ mode; ii) $a_{\rm self}$ reaches its peak value around $2.4 M$ and diverges at the horizon; iii) The lower panel shows that $\Delta r$ is positive, and at around $3.1$M, $\Delta r$ changes sign, possibly diverging at the horizon. Further, we discord on interpreting the phenomenon as purely repulsive: the ``radiation reaction effects [...] tend to decelerate the particle with respect to the zeroth order (Schwarzschild) geodesics. This is what one would qualitatively expect a priori
since the system is losing energy and momentum in the form of gravitational radiation";  the repulsive effect is confirmed in \cite{lo01}. Instead, we fully concord with the findings in \cite{balo02}, ``the radial component of the SF is found to point inward {\it i.e.}, toward the black hole throughout the entire plunge. [...] Consequently, the work done by the SF on the particle is positive, resulting in that the energy parameter E {\it increases} throughout the plunge." Among the findings in \cite{balo02}, we confirm also (i) the consistency of the mode-sum regularisation with the $\zeta$-function method; (ii) the values of the regularisation parameters; (iii) the plots of the self-force components $F^t$ and $F^r$; (iv) the asymptotic behaviour for large $\ell$.  
This is the first confirmation in the literature of the findings (i) - (iv). 
}.
Needless to say, the time shortness of the fall forbids any important accumulation of back-action effects but, from the epistemological point of view, radial fall for gravitation remains the most classical problem of all, and raising the most delicate technical questions. The dependence of the back-action on gauge is also not a valid argument to dismiss the eldest problem in physics, as gauge dependence is also present in the geodesic fall without back-action; it is inherent in general relativity, and even surfaces in Newtonian physics. Finally, the results herein are obtained (and coincide) in two gauges which are most used.  

In particle physics, namely transplanckian regime and black hole production, back-action has a pivotal role in head-on collisions
\cite{gakospto10b}.

The modern frame to study back-action is provided by the Extreme Mass Ratio Inspiral (EMRI) sources. Gravitational waves from compact stars or solar size black holes orbiting supermassive black holes are targeted by Space Laser Interferometry (SLI) \footnote{https://www.elisascience.org/}, now officially in the European Space Agency planning. The community aims to trace the most complex - but astrophysically plausible - orbits around rotating black holes. 
The last stages of EMRI plunge (a quasi-radial case) were analysed  for discriminating supermassive black holes from boson stars \cite{kegaka05}, being the latter horizonless objects, and  for signatures of dark matter \cite{mapacacr13}. 

Self-force refers to the MiSaTaQuWa-DeWh approaches  \cite{misata97,quwa97,dewh03} to compute the back-action, see \cite{blspwh11} for an introduction to the techniques for self-force computation. 

It is essential to recall that radial fall benefits of a regular gauge transformation \cite{baor01} between the harmonic (H) \footnote{The harmonic gauge was first used by Einstein \cite{ei16}, later by de Donder \cite{dd21}. It has also been attributed to Hilbert, Fock and Lanczos. Surely, it is not due to Lorenz who died when Einstein was two years old, or to Lorentz, co-father with Poincar\'e of the transformations bearing their names. Further, the Lorenz gauge in electromagnetism has been attributed to FitzGerald instead.} and the Regge-Wheeler (RW) \cite{rewh57} gauges.  

The quantities in \cite{lo00} were computed in the RW gauge, using the Riemann-Hurwitz $\zeta$ function \cite{ri59, hu82} (the $\zeta$ regularisation acts on the full perturbations $h_{\mu\nu}$, without a splitting of the  singular and regular parts). 
The metric is given by $g_{\mu\nu} + h_{\mu\nu}$, where $g_{\mu\nu}$ represents the background. The coordinate of the particle is $ 
{\hat r} = r + \Delta r$, where $\Delta r$ is the displacement due to the back-action. In the pragmatic method, the second order coordinate time derivative of the displacement is given by \cite{lo00,lo01, spao04, sp11} 

\beq
\Delta \ddot r =
a_1 \left(g_{\mu\nu}; r, {\dot r} \right)\Delta r 
+ a_2 \left(g_{\mu\nu}; r, {\dot r} \right) \Delta \dot r
+ a_{\rm self}\left(h_{\mu\nu}; r, {\dot r}\right ),
\label{leq}
\eeq
where the first two terms  form the background geodesic deviation (gd), while $a_{\rm self}$ is the perturbation dependent term (for their explicit expressions, see \cite{sp11}). 

Equation \ref{leq} 
corresponds to the first order deviation from geodesic motion (in proper time) found by Gralla and Wald \cite{grwa08, grwa11}, in the H gauge, Eq. \ref{gweq}. For an introduction see \cite{spriao14} in this journal. 
For $h_{\mu\nu}^R$ being the  tail or radiative part of the perturbations \cite{misata97, quwa97, dewh03}, the second order proper time covariant derivative is given by (${\hat x}^\alpha = x^\alpha + \Delta x^\alpha$)  \cite{grwa08,grwa11,spriao14} 
 
\beq
\frac{D^2 \Delta x^\alpha}{d\tau^2}=
\underbrace{- {R_{\mu\beta\nu}}^\alpha u^\mu \Delta x^\beta u^\nu}_{\rm Background~geodesic~deviation} 
\underbrace{- \frac{1}{2}
(g^{\alpha\beta} + u^\alpha u^\beta) 
(2h_{\mu\beta ;\nu}^R- h_{\mu\nu ;\beta}^R) u^\mu u^\nu}_{{\rm Self-acceleration~(proper~time)}~=~F^\alpha_{\rm self}/m}, 
\label{gweq}
\eeq
where $R_{\mu\beta\nu}^\alpha$ is the Riemann tensor and $u^\alpha$ the four-velocity.  
We have determined $a_{\rm self}$, Eq. \ref{leq}, via the Mode-Sum regularisation  \cite{baor00}. 

\begin{figure}[H]
\centering\includegraphics[width=10cm]{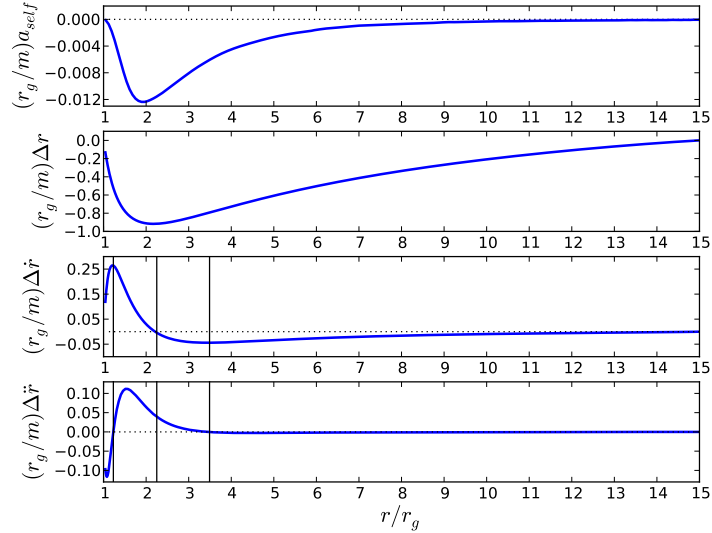}
\caption{The coordinate quantities $a_{\rm self}$, $\Delta r$, $\Delta {\dot r}$, and $\Delta {\ddot r}$ versus $r/r_{\rm g}$ for $r_0 = 15 r_{\rm g}$ (null initial velocity), in normalised and geometrised units. The values in SI units are obtained through multiplication by $m/r_{\rm g}$ - in geometrised units - and for $c^p$ ($p=0,1,2$ for length, velocity and acceleration, respectively). The vertical lines mark the four zones, Tab. \ref{tab1}.}
\label{fig1}
\end{figure}

In the H gauge the SF is obtained by subtracting the singular part of the force from the retarded force \cite{spriao14}

\beq
F^{\alpha(\text{H})}_\text{self}=F^{\alpha(\text{H})}_\text{ret}-F^{\alpha(\text{H})}_\text{S}~.
\label{ret-s}
\eeq

We recall the expression of the Mode-Sum decomposition in H gauge by Barack and Ori \cite{baor00}, and that the RW gauge is regularly connected to the H gauge for purely radial orbits \cite{baor01}. Furthermore, it has been shown that the components of the transformation gauge vector are not only regular at the position of the particle but they be made vanishing.  That is to say, the regularisation parameters share the same expression in RW and H  gauges. The SF is thus gauge invariant for RW, H and all other gauges interrelated via a regular transformation gauge vector. Further, the RW gauge has the distinct advantage of giving easy access to the components of the perturbation tensor, instead strongly coupled in the H gauge (see appendix).

From Eq. \ref{leq}, Fig. \ref{fig1}, we plot the coordinate quantities  $a_{\rm self}$, $\Delta r$, $\Delta {\dot r}$, and $\Delta {\ddot r}$ versus $r/r_{\rm g}$ for $r_0 = 15 r_{\rm g}$ and null initial velocity. The displacement $\Delta r$ is always negative, and four different zones can be identified, Tab \ref{tab1}.

\begin{figure}[H]
\centering\includegraphics[width=10cm]{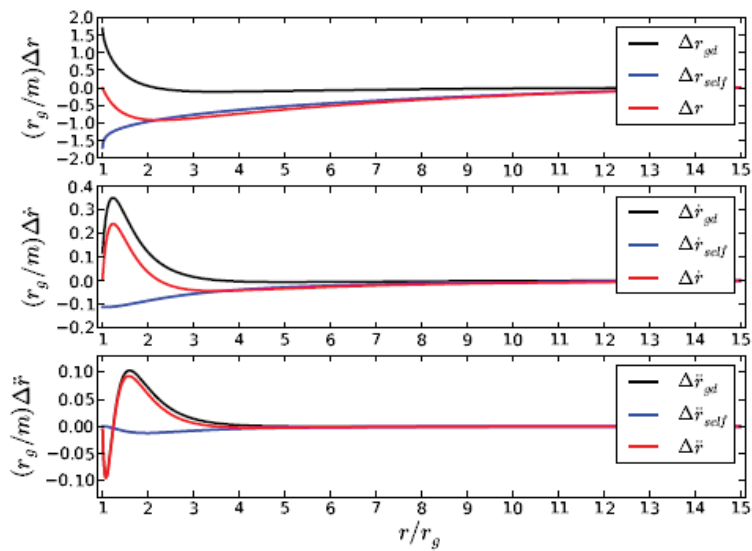}
\caption{Separate contributions by the geodesic deviation and the self-acceleration to the coordinate quantities of Fig. \ref{fig1}.}
\label{fig2x}
\end{figure}

\begin{table}[H]
\centering
\caption{The four zones according to the sign of $\Delta r$, $\Delta {\dot r}$, $\Delta {\ddot r}$. }
\label{tab1}
\begin{tabular}{@{}|l|c|c|c|c|}
\hline
Zone & IV & III & II & I \\
     & $r_{\rm g} \!-\! 1.2~r_{\rm g} $ 
     & $1.2~r_{\rm g}\!- \! 2.2~r_{\rm g}$ 
     & $2.2~r_{\rm g}\!- \! 3.5~r_{\rm g}$
     & $3.5~r_{\rm g}\!-\! r_0$ \\
\hline
$\Delta r$ & - & - & - & -\\
\hline
$\Delta {\dot r}$ & + & + & - & -\\
\hline
$\Delta {\ddot r}$ & -  & + & + & - \\
\hline
\end{tabular}
\end{table}

In zone I ($3.5~r_{\rm g} < r \leq r_0 = 15~r_{\rm g}$), the particle falls faster than in a background geodesic. Approaching the potential, it radiates more and it undergoes a breaking phase: in zone II ($2.2~r_{\rm g} < r < 3.5~r_{\rm g}$), the acceleration deviation $\Delta {\ddot r}$ becomes positive, but the velocity deviation $\Delta {\dot r}$ remains negative; in zone III ($1.2~r_{\rm g} < r < 2.2~r_{\rm g}$), the breaking is stronger and even the velocity deviation turns positive. Finally, in zone IV ($r_{\rm g} < r < 1.2~r_{\rm g}$), the acceleration deviation reappears negative, but not sufficiently to render the velocity deviation again negative.

\begin{figure}[H]
\centering\includegraphics[width=10cm]{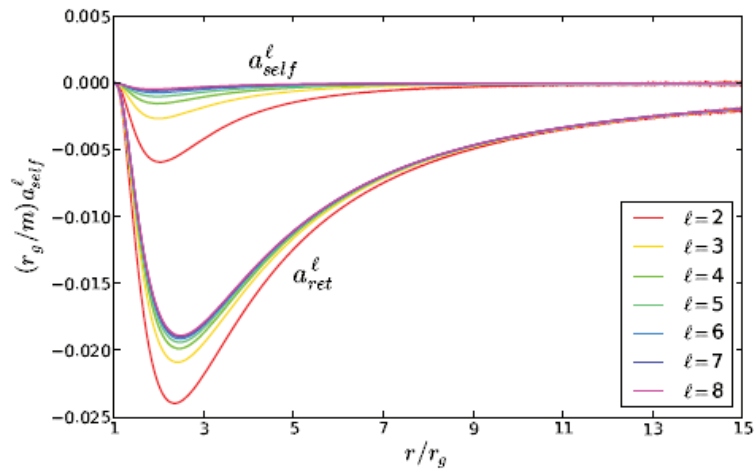}
\caption{Behaviour of $a_{\rm ret}$ and $a_{\rm self}$ {\it vis \`a vis} the ${\ell}$ modes.}
\label{fig3x}
\end{figure}

\begin{figure}[H]
\centering\includegraphics[width=10cm]{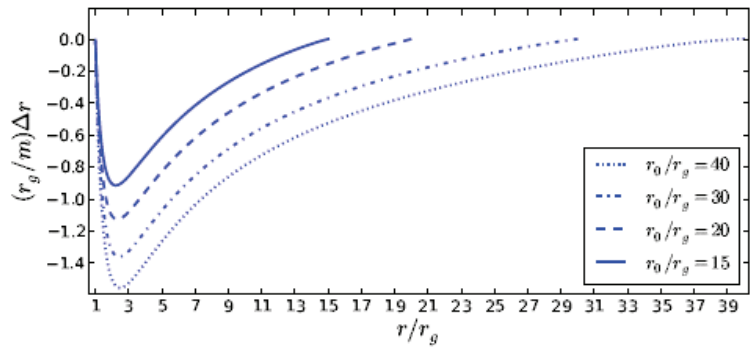}
\caption{The deviation $\Delta r$ for $r_0 = 15-40 r_{\rm g}$.}
\label{fig4x}
\end{figure}

Close to the horizon, all quantities approach zero, and thereby agree with the classic stand point of a far observer who doesn't see the particle crossing the horizon.      
In Fig. \ref{fig2x}, we plot the contributions to $\Delta r$, $\Delta {\dot r}$, and $\Delta {\ddot r}$ by 
$\Delta {\ddot r}_{\rm gd}$, $\Delta {\ddot r}_{\rm self}$. The contributions are obtained through a system of coupled equations $\Delta {\ddot r}_{\rm gd} = a_1\Delta r + a_2 \Delta {\dot r}$ and $\Delta {\ddot r}_{\rm self} = a_{\rm self}$, where $\Delta r = \Delta r_{\rm gd}+ \Delta r_{\rm self}$. 
The self-quantities act always inward the black hole. Conversely, the geodesic deviation is repulsive (except the acceleration close to the horizon), and often of larger magnitude than self-quantities. Incidentally, $a_1\Delta r$ and $a_2\Delta {\dot r}$ have opposite signs. 
The maximal coordinate velocity $|dr/dt|_{\rm max}$ is not any longer $0.3592c$ but increases of $5 m/M \% $, while $r_{\rm max} = 2.647 r_{\rm g}$ moves towards larger r. 
There is an uniform behaviour of $a_{\rm ret}$ and $a_{\rm self}$ {\it vis \`a vis} the different ${\ell}$ modes, and there is no diverging behaviour at the horizon, Fig. \ref{fig3x}. The back-action is dominated by the $l = 2$ mode, about 55\% of the total. 
Increasing $r_0$, Fig.  \ref{fig4x}, $\Delta r$ increases (like $\Delta {\dot r}$).

\begin{figure}[H]
\centering\includegraphics[width=8cm]{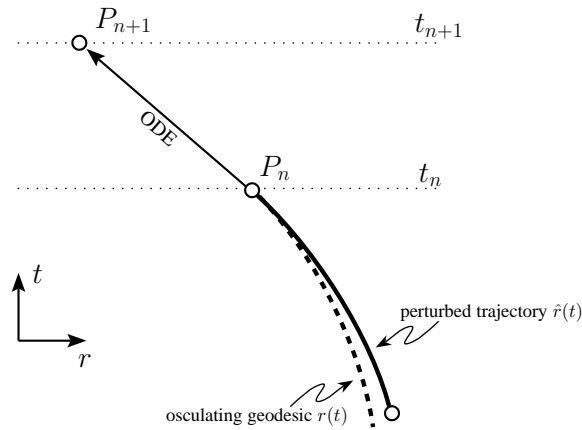}
\caption{The self-consistent (osculating) scheme.}
\label{fig5}
\end{figure}

\section{The self-consistent method}

The adiabatic approximation requires that a given orbital parameter changes slowly when compared to the orbital time-scale. The above condition is not satisfied for radial fall \cite{quwa99}. As we have seen, the feebleness of cumulative effects does not imply the non-existence of the self-force, and the lack of adiabaticity obliges to add the uppermost care in the computation. 

Gralla and Wald consider than any perturbation scheme is doomed to failure at late times, and suggest to evolve the most relativistic orbits through the iterative application of the back-action on the particle world-line, {\it id est}, the self-consistent approach \cite{grwa08, grwa11}. We implement it for the least adiabatic orbit of all, that is radial infall, and for the first time. It is worth stating that the strict self-consistency implies that the applied self-force at some instant is what arises from the actual field at that same instant. 
So far this has been done only for a scalar charged particle around an SD black hole \cite{divewade12}, and never for a massive particle. 
In other works \cite{waakbagasa12,labu12}, the applied self-force is what would have resulted if the particle
were moving along the geodesic that only instantaneously matches the true orbit. Herein, we adopt the latter acception. 
Our approach in orbital evolution (in RW gauge) consists thus in computing the total acceleration through self-consistent (osculating) geodesic stretches of orbits of iterative index $n$ (Fig. \ref{fig5}) 

\begin{figure}[H]
\centering\includegraphics[width=11cm]{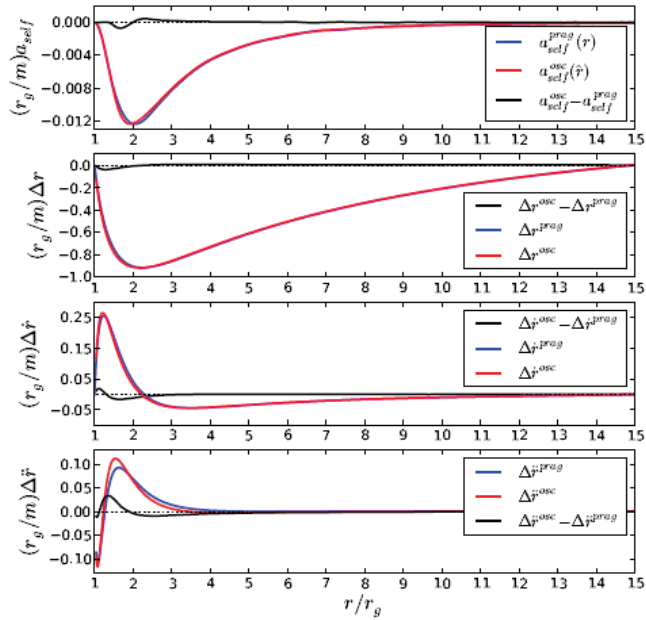}
\caption{$a_{\rm self}$, $\Delta r$, $\Delta {\dot r}$, and $\Delta {\ddot r}$ from the pragmatic and self-consistent (osculating) methods, and their difference. }
\label{fig6x}
\end{figure}

\beq
\ddot {\hat r} =
a_0\left(g_{\mu\nu}; {\hat r}, {\dot {\hat r}}\right) + a_{\rm self}\left(h_{\mu\nu}; {\hat r}, {\dot {\hat r}}\right), 
\label{sceq}
\eeq
where $a_{\rm self}$ is the (coordinate) self-acceleration at first order
\[
a_0 = - \frac{1}{2} g_{00}({\hat r})\frac{dg_{00}({\hat r})}{d {\hat r}}\left[ 1 - \frac{3{\dot {\hat r}}^2}{g_{00}({\hat r})^2}\right]~.
\]

At each integration time-step $t=t_n$ and at the $P_n$ point, a geodesic $r_n(t)$ osculating the perturbed trajectory $\hat{r}_n(t)$ is searched. The identification of a new ($t_0$, $r_0$) starting point is sufficient to determine such geodesic, that passes at $P_n$ at the right time and velocity. The self-acceleration is then computed along the geodesic until the point $P_n$. The solution of the ordinary differential equation, Eq. \ref{sceq}, leads to the new $P_{n+1}$ point.

In Fig. \ref{fig6x}, for  $z_0 = 15 r_{\rm g}$, we show (pragmatic and self-consistent methods)   
$a_{\rm self}$, $\Delta r$, $\Delta {\dot r}$, and $\Delta {\ddot r}$. The amplitude of these quantities, when computed self-consistently, differ of about $3\%$ after $4 r_{\rm g}$, and the four zones are slightly shifted towards the horizon. In our simulation, differences between the pragmatic and self-consistent (osculating) methods appear in the values of motion related quantities. 
 
\section{Conclusions and perspectives}

In the Regge-Wheeler and de Donder (harmonic) gauges and all other smoothly related gauges, a far observer concludes that the self-force pushes inward the falling body, with a strength proportional to the mass of the small body for a
given large mass. Further, the value of the maximal coordinate velocity raises, this value being reached earlier on during infall. When a self-consistent approach is adopted, and thereby the trajectory is iteratively corrected by the self-force, these effects are further emphasised. 

We have clarified the relation of the back-action with radial fall, also in terms of epistemological significance. For future developments, the self-consistent method appears a truly innovative technique to tackle the complex highly relativistic and non-adiabatic orbits that compact objects trace around supermassive black holes.  

\section*{Acknowledgments}

Discussions with S. Aoudia (Golm), S. Cordier and S. Jubertie (Orl\'eans) are acknowledged.  

\section*{Appendix}

For $F^{r,t}_{\rm self}$, radial and time components of the self-force ({\rm "ret"} the retarded field; $\ell$ the (polar) mode, $L = \ell + 1/2$;  A,B,C,D the regularisation parameters computed in RW gauge), we have

\beq
F^\alpha_{\rm self}= \sum_{\ell = 0}^\infty\left[F^{\alpha\ell}_{\rm ret \pm} - A^\alpha_\pm - B^\alpha - 
C^\alpha L^{-1}\right] - D^\alpha;
\eeq
after averaging, ($B_a = {\ds \frac{g_{00}}{m g_{00}(r_0)}}[B^r - {\dot r} B^t]$), the (coordinate) self-acceleration is 

\beq
a_{\rm self} = 
\frac{ g_{00}^2}{m g_{00}(r_0)}(F^r_{\rm self} - 
{\dot r} F^t_{\rm self}) = \sum_{\ell = 0}^{\infty}\left[a_{\rm ret}^{\ell} - B_a \right] 
= a_{\rm self}^{\ell = 0} + \sum_{\ell = 2}^{\ell_{\rm max}}a_{\rm self}^{\ell} + 
\sum_{\ell_{\rm max} + 1}^{\infty}a_{\rm self}
^{\ell \rightarrow \infty }.
\eeq  

With a proper gauge choice, the $\ell =1$ - polar mode - vanishes. The second and third terms are numerically and analytically obtained, respectively (the latter through the polygamma function, related to the $\zeta$ function \cite{rietal14}; 
error $\leq 0.1 \%$ for $\ell_{\rm max} = 8$).

\bibliography{references_spallicci_140721}

\end{document}